\documentclass[letterpaper,10pt,oneside,journal,final]{IEEEtran}
\usepackage{amsmath}
\usepackage{amssymb}
\usepackage{amsfonts}
\usepackage{graphicx}
\usepackage{array}
\usepackage{booktabs}
\usepackage{floatrow}
\usepackage{cite}
\usepackage{multirow}
\usepackage{balance}
\usepackage{acronym}
\usepackage[binary-units = true]{siunitx}
\usepackage{tikz}
\usepackage{textcomp}
\usepackage{hyperref}
\usepackage{ccicons}

\acrodef{res}[\textsc{res}]{residential}
\acrodef{nrs}[\textsc{nrs}]{non-residential}
\acrodef{pvg}[\textsc{pvg}]{photovoltaic}
\acrodef{plt}[\textsc{plt}]{public lightning}
\acrodef{mix}[\textsc{mix}]{mixed}
\acrodef{cty}[\textsc{cty}]{city}

\IEEEoverridecommandlockouts

\newcommand\copyrighttext{%
  \footnotesize
  \centering\ccbyncnd\\ This work is licensed under a Creative Commons Attribution-NonCommercial-NoDerivatives 4.0 International License.\\
  DOI: \href{https://doi.org/10.1016/j.segan.2020.100308}{10.1016/j.segan.2020.100308}}
\newcommand\copyrightnotice{%
\begin{tikzpicture}[remember picture,overlay]
\node[anchor=south,yshift=0pt] at (current page.south) {\setlength{\fboxrule}{0pt}\fbox{\parbox{\dimexpr\textwidth-\fboxsep-\fboxrule\relax}{\copyrighttext}}};
\end{tikzpicture}%
}

\begin{document}
\title{Functional Principal Component Analysis as a Versatile Technique to Understand and Predict the Electric Consumption Patterns}

\author{Davide~Beretta, Samuele~Grillo, Davide~Pigoli, Enea~Bionda, Claudio~Bossi, and Carlo~Tornelli%
\thanks{This work has been financed by the Research Fund for the Italian Electrical System in compliance with the Decree of April 16, 2018.}%
\thanks{D. Beretta was with RSE SpA, via R. Rubattino, 54, I-20134 Milano (MI), Italy. (e-mail: davide.beretta@empa.ch)}%
\thanks{D. Pigoli is with the Department of Mathematics, King's College, London, WC2R 2LS, UK (e-mail: davide.pigoli@kcl.ac.uk).}%
\thanks{S. Grillo, is with the Dipartimento di Elettronica, Informazione e Bioingegneria, Politecnico di Milano, piazza Leonardo da Vinci, 32, I-20133 Milano, Italy (e-mail: samuele.grillo@polimi.it).}%
\thanks{E. Bionda, C. Bossi and C. Tornelli are with RSE SpA, via R. Rubattino, 54, I-20134 Milano (MI), Italy (e-mail: \{enea.bionda, claudio.bossi, carlo.tornelli\}@rse-web.it).}}

\IEEEaftertitletext{\copyrightnotice\vspace{1.1\baselineskip}}
\maketitle

\begin{abstract}

Understanding and predicting the electric consumption patterns in the short-, mid- and long-term, at the distribution and transmission level, is a fundamental asset for smart grids infrastructure planning, dynamic network reconfiguration, dynamic energy pricing and savings, and thus energy efficiency. This work introduces the Functional Principal Component Analysis (FPCA) as a versatile method to both investigate and predict, at different level of spatial aggregation, the consumption patterns. The method was applied to a unique and sensitive dataset that includes electric consumption and contractual information of Milan metropolitan area. The decomposition of the load patterns into principal functions was found to be a powerful method to identify the physical and behavioral causes underlying the daily consumptions, given knowledge of exogenous variables such as calendar and meteorological data. The effectiveness of long-term predictions based on principal functions was proved on Milan's metropolitan area data and assessed on a publicly-available dataset.

\end{abstract}



\acresetall
\section{Introduction}\label{sec:Introduction}
\IEEEPARstart{T}{he} ability to predict the daily electric consumption, both at the distribution and transmission level, in the short-, mid- and long-term, is an essential asset in the evolving scenario of modern cities, where the recent advances in the microelectronics and in the information and communication technologies are finally making possible the transition from the old centralized energy management model to the smart grid paradigm, ever more characterized by the presence of energy generation from distributed sources~\cite{meadows2012limits,greengard2015internet,zhou2016big,mayer2013big}. The advantages coming from the prediction of the electric consumption, and the corresponding daily load patterns, are manifold and include, at different levels of spatial aggregation, energy savings, infrastructure planning, and energy pricing~\cite{chan2012load,khan2016load}. On the one hand, the short term prediction at the distribution level can be exploited to re-configure the electric network on a timely manner, increasing or decreasing the security level of restricted network regions on the basis of specific needs, and to provide basic information for more customer-based personalized market policies and electric power services, with the combined ultimate scope of preventing load peaks. In the same time window, the prediction at the transmission level can provide useful information to regulate the energy generation and provisioning to macro areas, ultimately allowing to reduce the waste of non-storable electric energy and thus apply favorable energy pricing. On the other hand, the mid- and long-term prediction is beneficial, at both the distribution and transmission levels, to operations planning and infrastructures improvements, with an expected stronger impact when the consumption forecasts are combined with projections on population. In this context, analysis methods that aim at unveiling the physical causes underlying the consumption patterns at different levels of spatial aggregation are of unparalleled importance to gain insights into the mechanisms regulating the electric loads.
A variety of different techniques and methodologies has been so far proposed, unraveling the dependency of the daily load patterns on single customers' behavior as a function of the spatial aggregation~\cite{capasso1994bottom,walker1985residential,henley1997non,valor2001daily,pardo2002temperature,christenson2006climate,engle1992modelling,hernandez2012study}.
Among those techniques, the analysis and the classification are generally pursued by means of time-series clustering methods~\cite{hernandez2012study,linden2017categorisation,rhodes2014clustering,panapakidis2012electricity,rasanen2010data,chicco2012overview}, which suffer from high sensitivity to the choice of the metric~\cite{liao2005clustering}, while the predictions are based on parametric and non-parametric methods~\cite{suganthi2012energy}, such as multi-regression and auto-regressive historical time series~\cite{valor2001daily,pardo2002temperature,engle1992modelling,hosking2013short,massana2015short,moral2005modelling}, and  exponential smoothing~\cite{gardner1985exponential,christiaanse1971short,taylor2003short}. In some cases, the prediction follows data dependencies reduction by means of principal component analysis. In all of the aforementioned methods, the physical causes underlying the load patterns remain undisclosed, as the techniques, regardless of being more or less effective, are based on mathematical tools that are transparent to the shape of the analyzed patterns.
This work introduces the use of the Functional Principal Component Analysis (FPCA)~\cite{ramsay} as an alternative and effective method (i) to investigate the daily electric load patterns at different level of aggregations, with a spatial-aggregation-dependent level of accuracy, providing an unparalleled way to correlate the observed consumption patterns with exogenous causes, and (ii) to predict the daily electric consumption patterns in the short- and long-term.
The main contribution of this work is therefore twofold. On the one hand, it shows that the FPCA can be used as a tool to analyze complex dataset of electric consumptions\footnote{In our case, we applied the proposed methodology to a dataset which gathers consumption measurements from approximately 3500 MV/LV secondary substations for three years with a \SI{15}{\minute} sampling time. This amount of data corresponds to a 350-million-tuples table.} and to understand the physical and behavioral causes underlying the load patterns. On the other hand, it shows that the FPCA decomposition can be exploited to predict the electric consumptions in the long-term, with very competitive performances.

\section{Methods}\label{sec:SystemModel}
\subsection{Functional Principal Component Analysis (FPCA)}

The FPCA is a statistical method to represent functional data in an orthonormal basis of the Hilbert space that consists of the eigenfunctions of the covariance operator. If $f_i\left(t\right)$ is the $i$-th functional data, or curve, of the variable $t$, according to the FPCA~\cite{ramsay}
\begin{equation}
  f_i\left(t\right) = \mu\left(t\right) + \sum_{k=1}^{p} c_{k,i}\varphi_k\left(t\right),
\end{equation}
where $\mu\left(t\right) = n^{-1}\sum_{i=1}^{n}f_i\left(t\right)$ is the average of the input functions and $\varphi_k\left(t\right)$ are the eigenfunctions of the sample covariance operator
\begin{equation}
  S\left(t,\tau\right) = \frac{1}{n-1}\sum_{i=1}^{n}\left( f_i\left(t\right)-\mu\left(t\right)\right)\left( f_i\left(\tau\right)-\mu\left(t\right)\right),
\end{equation}
defined such that $\langle  S\left(t,\tau\right)\varphi_k\left(\tau\right)|\varphi_k\left(t\right)\rangle = \lambda_k$, being $\langle\cdot | \cdot \rangle$ the inner product and $\lambda_k$ the $k$-th eigenvalue of $S$. The $\varphi_k\left(t\right)$ are usually called functional principal components and the $c_{k,i}$, which are defined with sign, are given the name of scores. The shape of the $\varphi_k\left(t\right)$, together with the sign of the scores, determines the ``direction'' of variability of the data with respect to the average consumption pattern $\mu\left(t\right)$. The absolute value of the score gives information about how much the consumptions deviate from the average. The scores vector $c_{k,i}$,\ with $k \in[1,p]$, gives the representation of the $i$-th curve in the new basis. The principal component basis is ordered in the sense that $\sum_{i}^{n}{\left\langle f_i\left(t\right)-\mu\left(t\right)\middle|\varphi_k\left(t\right)\right\rangle}^2>\ \sum_{i}^{n}{\left\langle f_i\left(t\right)-\mu\left(t\right)\middle|\varphi_l\left(t\right)\right\rangle}^2$, for every $l > k$. This means that the first $p$ functional principal components provide the best possible approximation (in a square error sense) of the data among all the possible sets of basis functions with $p$ elements. Moreover, by definition of the functional principal components, the variance of the (centered) data projected in the direction of the $k$-th component is the $k$-th eigenvalue of the sample covariance operator $S$. This peculiarity can be exploited to select a number of principal components sufficient to represent the functional data with a certain level of precision, e.g. more than 90\% of the variability of the data, or such that the addition of further components does not improve significantly the amount of information reproduced. The cumulative proportion of variability explained by the first $p$ principal components can be computed as $\vartheta\left( p \right)=\ \sum_{i}^{p}{\lambda_i/\sum_{i}^{n}\lambda_i}$. In this work, the $f_i\left(t\right)$ is the daily electric consumption pattern of the $i$-th day of a given electric station or spatial aggregation of electric stations, where $t \in [0,23]$ is the discrete time instant, i.e. the hour of the day, at which $f_i\left(t\right)$ is evaluated. The FPCA was exploited to identify correlations between the shape of the $\varphi_k\left(t\right)$ and exogenous variables such as the day of the week, the month of the year, the temperature, and the relative humidity, to name a few. This part of the analysis was not trivial and required a graphical representation of the components and of the scores as a function of the exogenous variables to visualize trends and behaviors, besides a decent knowledge of the conditions under which the consumption patterns were generated. To the purpose, the present study was extensively supported by graphs showing: i) the first three $\varphi_k\left(t\right)$, which were demonstrate to explain more than the 80\% of the variability of each consumption pattern analyzed, ii) the daily and monthly distribution of the scores, and iii) the distribution of the scores as a function of the temperature and of the relative humidity. The last figure had particular relevance in the study as it allowed to identify a comfort zone that corresponds to a minimum in the electric consumption.

\subsection{Prediction}
The FPCA can be integrated into any time-series predictive model to predict future patterns. The peculiarity of  the FPCA-based predictive models is that they predict the future values of the $c_{k,i}$ and thus, since the $\varphi_k\left(t\right)$ are orthogonal to each other, they can predict the scores of a selected subset of the $\varphi_k\left(t\right)$ without necessarily calculate all the $c_{k,i}$ simultaneously. This property can be exploited to reach a compromise between the complexity of the model and the explained variability predicted.
In this framework, the chosen generic predictive model was linear, i.e.,
\begin{equation}
  c_{i,k} = \boldsymbol{x}^{T}_{i,k}\boldsymbol{\beta}_{k} + \varepsilon_{i,k}
\end{equation}
where $\boldsymbol{x}_{i,k}$ is the vector of the predictors for the $k$-th FPC score of the $i$-th day, $\boldsymbol{\beta}_{k}$  is the vector of the coefficients (to be estimated) associated to the $k$-th score, and $\varepsilon_{i,k}$ are the zero-mean independent Gaussian errors. The coefficients are estimated via ordinary least squares and the estimated model is used to predict the future scores.
The set of predictors used for each score was dynamically selected with a stepwise procedure in order to minimize the Akaike Information Criterion (AIC), i.e.,
\begin{equation}
  {\rm AIC} = 2p-2\ln L,
\end{equation}
where $p$ is the number of the predictors of the model and $L$ is the corresponding likelihood function. This approach allowed to choose a model showing a good compromise between complexity and accuracy. Since the error model was assumed Gaussian, the likelihood function was proportional to the sum of the square of the residuals. The pool of predictors included the calendar time (number of days passed from the first recorded observation, to account for long term trends), the month of the year (categorical variable represented by eleven indicator variables, to account for seasonal effects), the day of the month (continuous variable to account for inter-month trends), the day of the week (categorical variable represented by six indicator variables, to account for working/festive patterns), and the presence of popular events (three indicator variables to signal the presence of Milan Fashion Week, Expo 2015, Milan Design Festival), which might affect the consumption patterns due to the abnormal presence of people in specific geographical areas. The meteorological variables can be included for short-term predictions in case accurate forecasting or other proxies are made available.

In this work, the analysis focused on two different predictions, i.e., i) the daily consumption pattern, and ii) the monthly average energy consumption, which belongs to short- and mid/long-term prediction cases, respectively. The goodness of the prediction of the daily consumption patterns was measured by the Mean Absolute Percentage Error (MAPE), that is
\begin{equation}
  {\rm MAPE} = \frac{1}{m}\sum_{i=1}^{m}\left | \frac{x_i - y_i}{x_i}\right | \times 100,
\end{equation}
where $x_i$ and $y_i$ are the measured and predicted values of the test data set, respectively, and $m$ is the number of predicted values that define the time series in a specific time window. In this context, $m = 24$ is the number of samples that define the daily consumption pattern of a given day of the year. The goodness of the prediction on the monthly average energy consumption was measured by the energy percentage error $\varepsilon_{\%}$, that is
\begin{equation}
  \varepsilon_{\%} = \frac{1}{m} \frac{\left | \sum_{i=1}^{m}x_i - \sum_{i=1}^{m}y_i \right |}{\sum_{i=1}^{m}x_i} \times 100,
\end{equation}
where $x_i$ and $y_i$, and $m$ are again the measured and predicted values, and the number of samples, respectively.

\section{Data}
The initial data set comprised: i) the measurements, taken every \SI{15}{\minute}, of the average power of 3386 electric secondary MV/LV substations deployed on Milan's metropolitan area; ii) the geo-localization of the substations; iii) the contractual information of the customers connected to 5312 secondary substations deployed on Milan's metropolitan area, including both consumption and generation, and reporting, for each contract, start and expiry date, allocated maximum power and type of customer, i.e., residential, non-residential or public lighting; iv) the meteorological data, including temperature, relative humidity, radiation, rainfall and wind speed, recorded every hour by a number of ARPA's\footnote{ARPA is the ``Agenzia Regionale per la Protezione Ambientale'', i.e., Regional Agency for Environmental Protection.} meteorological stations deployed on Milan's metropolitan area; v) the calendar of national and local festivity, including particular events characterized by massive affluence, i.e., the fashion week, the design week and EXPO 2015; and vi) the geographical borders of Milan's neighborhoods. Data i)--iii) were made available by Milan's Distribution System Operator (Unareti S.p.A.), while data iv) and vi) were freely downloaded from ARPA and Milan's municipality website, respectively. Data v) was generated on the basis of the information freely retrieved from the web. Data covered, almost completely, the period from January 2014 to October 2017, and required some elaboration to take into account for reading failures and for different averaging time windows, local timezone and daylight saving time. Since the ARPA's meteorological stations deployed on Milan's area were non-uniformly distributed, the weather conditions were averaged over the values measured by all the stations. Data on customers were aggregated to the substation level in order to produce a large database, which could be easily queried, reporting the history of each substation. Considered the volume of the available data (tens of \si{\giga\byte}), within the framework of an ever increasing and potentially auto-updating dataset, data storage, elaboration and interrogation were approached following big data methodologies, using the Apache Spark framework, and exploiting the Amazon Web Services Simple Cloud Computing and Simple Cloud Storage Service.
The study was limited to the population of electric stations showing stable contractual characteristics in the period 2014--2017.  A given substation was considered stable with respect to a particular contractual characteristic $x$, in a specific time window, if the characteristic $x$, resulting from the aggregation over all the customers connected to that particular substation, satisfied the condition
\begin{equation}
  \max\left(x\right) - \min\left(x\right) <0.1 \operatorname{avg}\left(x\right),
\end{equation}
in the time interval considered. In this study, the contractual characteristics considered were: i) contractual consumption, ii) fraction of contractual consumption absorbed by residential customers, iii) fraction of contractual consumption absorbed by public lighting, iv) contractual generation, and v) fraction of contractual generation supplied by photovoltaics. The result of the operation applied to the database for the time interval 2014--2017, returned 3642 stations out of 5312, i.e., approximately 68\% of the initial population of substations the contractual characteristics of which were made available by the local electric distribution provider. The population of stable electric stations was used to build the non-normalized probability distribution functions (PDFs) and the normalized cumulative distribution functions (CDFs) of a subset of contractual characteristics, with the aim to visualize the framework of the substations deployed on Milan's metropolitan area from the point of view of the power supplier company. This allowed to define a window for each of the contractual characteristics that takes into account for the most representative substations of Milan's metropolitan area, and to identify the presence of substations showing particular contractual characteristics, the analysis of which could explain specific load patterns. In particular, the PDFs revealed the existence of a number of substations having contractual consumption (almost) entirely absorbed by residential, non-residential, or public lighting customers. Most of the electric stations display aggregated contractual consumption in the interval 500--1500 \si{\kilo\watt} and fraction of consumption due to residential customers in the range 30--70\%. Data also reveal that the fraction of consumption absorbed by public lighting is less than 1\% of the contractual consumption, that the contractual generation is always less than \SI{10}{\kilo\watt}, and that the contractual generation is entirely provided by photovoltaics.
Since the power measurements provided by the DSO covered only a subset of the stable electric stations registered in the customers contracts database, the FPCA of the load patterns was limited to the power curves of a subset of 2523 electric stations out of the 3386 initially available. Different level of spatial aggregations were considered, i.e., the single electric station, the neighborhood and the whole Milan's city area. To this purpose, the data on the electric stations were aggregated to the \textsc{nil} (Nucleo di Identit\`a Locale, i.e. neighborhood) and Milan levels, respectively. In each case, a database with the unique identification code of the entity (substation, \textsc{nil} or Milan), the aggregated power measurements, the date and day time of the recorded events, the spatially-averaged weather conditions and the aggregated contractual characteristics of the entity was built, allowing again to easily access the aggregated data by performing simple database queries. As the FPCA aimed at understanding the complexity of the electric load patterns by unveiling the exogenous causes that underlie the consumptions, a series of filters were applied to the set of electric stations to extract a number of populations of stations showing specific contractual characteristics, stable in the time interval 2014--2017. These populations were given the name of \ac{res}, \ac{nrs}, \ac{plt}, \ac{pvg}, \ac{mix}, \textsc{nil} and \ac{cty}, respectively, and the criteria used to select them are reported in Table~\ref{tab:tab_1}. A filtering process was necessary to remove all the stations affected by incomplete and/or corrupted data. This process followed three distinct steps: i) deletion of days characterized by incomplete data, i.e., with less than 24 entries per day per electric station. The substations having a number of incomplete days higher than 20\% of the available calendar days were removed completely from the sub-populations; ii) deletion of days characterized by corrupted data. The \ac{res}, \ac{nrs} and \ac{pvg} were considered corrupted if the power reading was equal to \SI{0}{\kilo\watt} at least once within the day. The sub-population of \ac{plt} was considered corrupted if the power reading was equal to \SI{0}{\kilo\watt} for the whole day. The corresponding substations were deleted from the sub-populations if the number of corrupted days was higher that 10\% of the available calendar days; iii) deletion of the substations showing non-corrupted data in less than 1095 days (approximately three years of data). The number of stations belonging to each population after the filtering is reported in Table~\ref{tab:tab_1} under the column ``FPCA''. They represent the final dataset over which the FPCA described in the following was applied to.

\begin{table}[ht]
	\centering
    \caption{List of populations considered in this work, along with their contractual consumption and characteristics, and their numerosity.}\label{tab:tab_1}
    \includegraphics[width=.8\columnwidth]{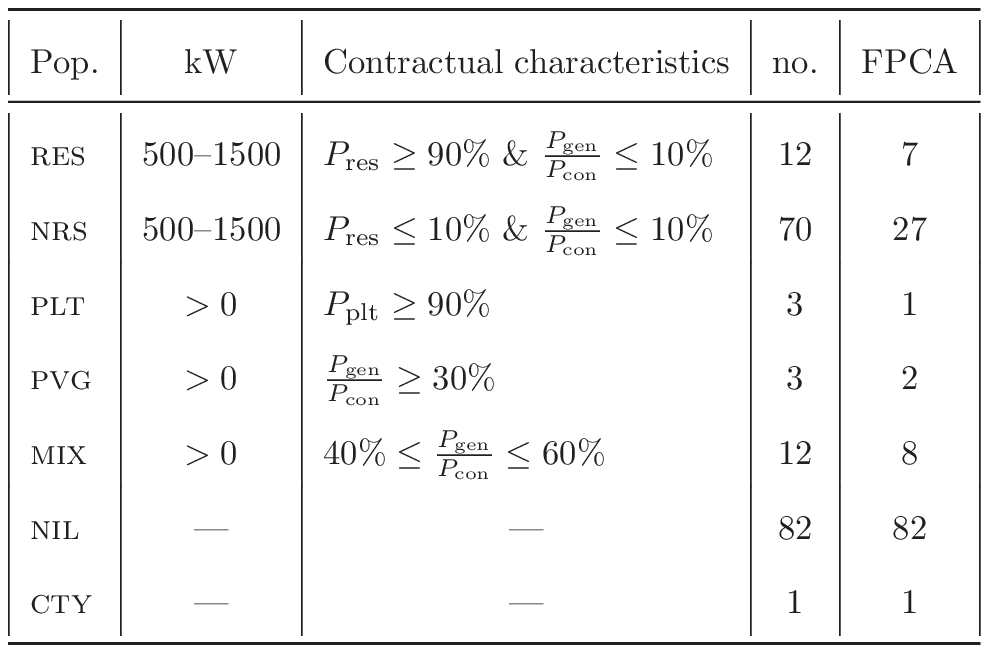}
\end{table}

\section{Result}
\subsection{Load pattern analysis}
\begin{figure}[t]
	\centering
    \includegraphics[width=.7\columnwidth]{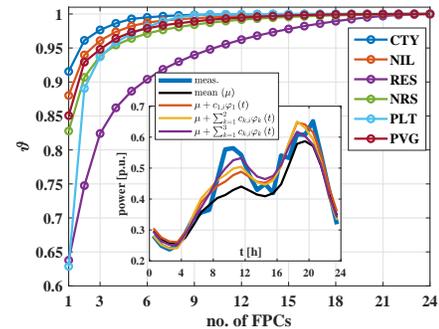}
    	\caption{Cumulative proportion of explained variability $\vartheta$, averaged over the population set, per population and plotted as a function of the ordered number of functional principal component. The inset plot shows the effect of progressively adding each FPC with the corresponding score to the mean.}
	\label{fig:fig_1}
\end{figure}

\begin{figure*}[t]
	\centering
    \includegraphics[width=.95\textwidth]{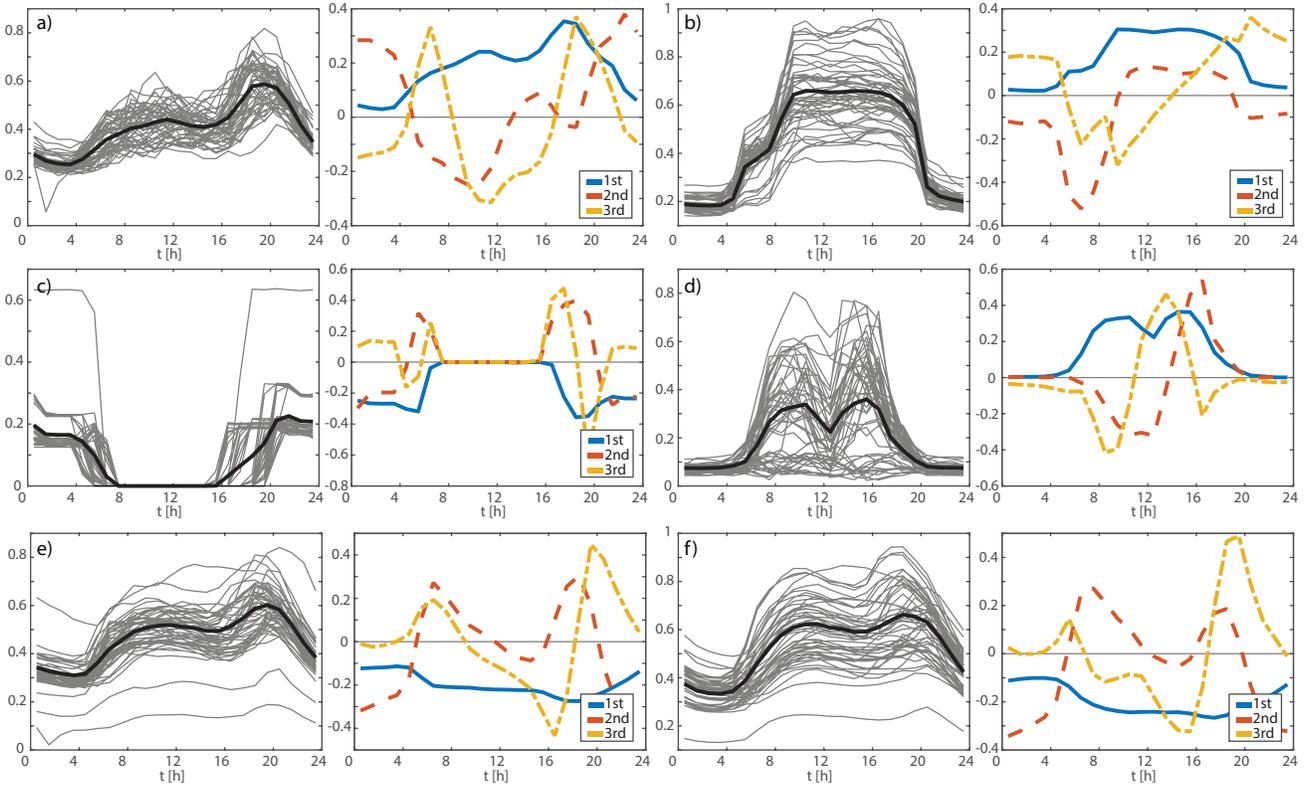}
    	\caption{The panel figure shows, on the left hand-side subplots, the daily average power load $\mu\left( t \right)$ superimposed on sample daily power load patterns and, on the right hand-side subplots, the first three FPCs (i.e., $\varphi_1\left(t\right)$ solid blue line, $\varphi_2\left(t\right)$ dashed red line, and $\varphi_3\left(t\right)$ dashed yellow line) of the analyzed elements taken from each population: a) \ac{res}, b) \ac{nrs}, c) \ac{plt}, d) \ac{pvg}, e) \textsc{nil}, and f) \ac{cty}.}
	\label{fig:panel1}
\end{figure*}

Figure~\ref{fig:fig_1} shows $\vartheta$ as a function of the number of $\varphi\left(t\right)$ for each population, together with an example of pattern reconstruction as a function of the number of $\varphi\left(t\right)$. $\vartheta$ increases with the spatial aggregation level, and the lowest $\vartheta$ is observed for the \ac{res} population, which is subjected to more variability and is thus harder to reproduce with a small number of $\varphi\left(t\right)$. Since, on average, more than 80\% of the variability of each population can be explained by the first three $\varphi\left(t\right)$, the study was limited to the first three $\varphi\left(t\right)$ only. The amount of data generated is cumbersome. Therefore, the analysis that follows is limited to a single element per population (see Table~\ref{tab:tab_1}), herein denoted with the same name of the population it belongs to, exception made for the population \textsc{mix}, which is not here reported.
The patterns were individually normalized with respect to the maximum power measurement displayed by each of them in the entire data set, i.e., the years 2014--2018. The normalization was crucial as it allowed for a direct comparison between the values of the scores of each $\varphi_k\left(t\right)$, even if belonging to different electric stations or level of spatial aggregation. The $f_i\left(t\right)$ and $\mu\left(t\right)$, together with the first three $\varphi\left(t\right)$, and the effect of the scores of the first three $\varphi\left(t\right)$ on $\mu\left(t\right)$, are grouped by population and shown in a series of sub-panels in Figure~\ref{fig:panel1}.
\begin{figure*}[t]
	\centering
    \includegraphics[width=.95\textwidth]{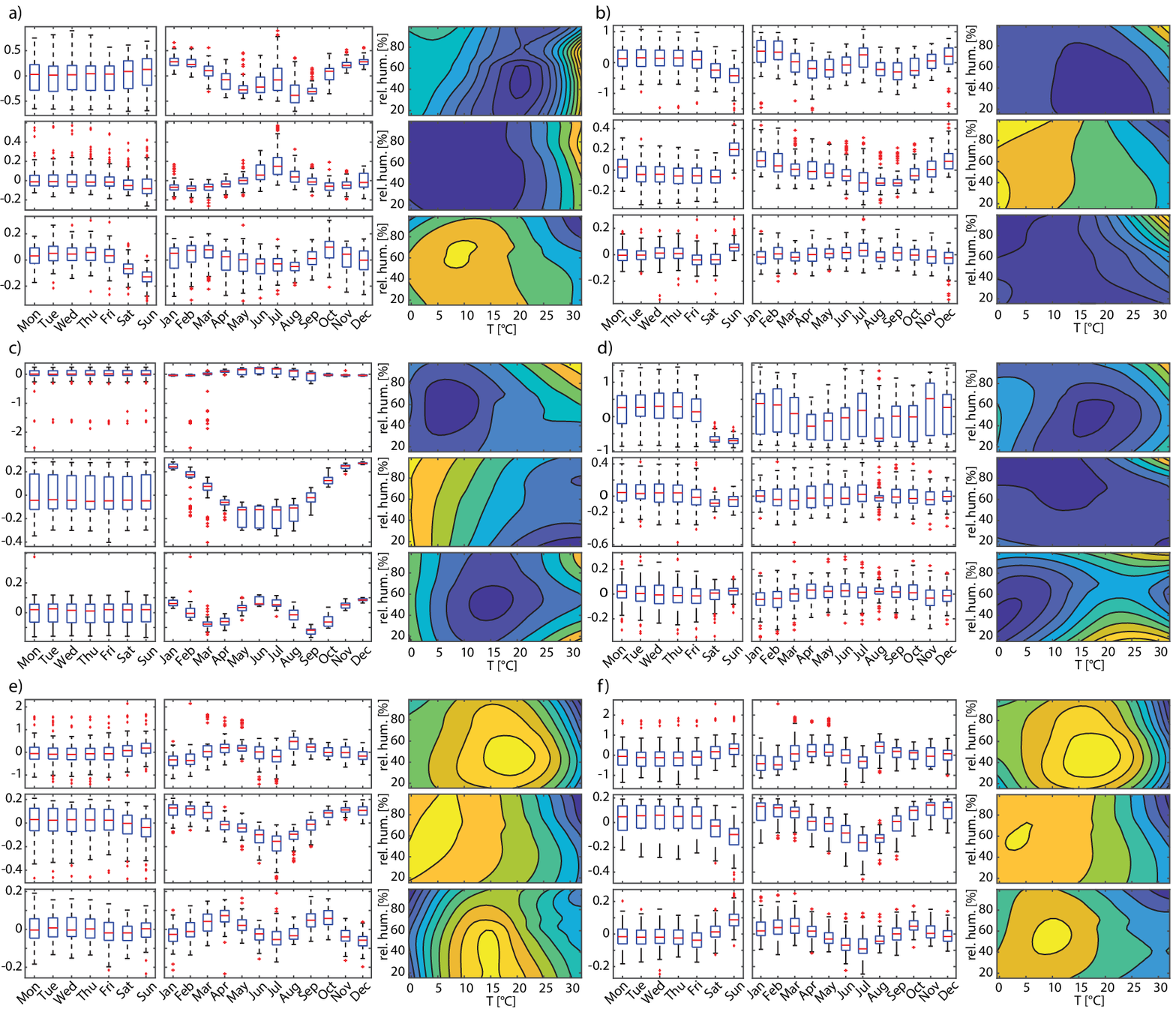}
    	\caption{Weekly, monthly and weather-dependent trends of the scores of representative elements belonging to the a) \ac{res}, b) \ac{nrs}, c) \ac{plt}, d) \ac{pvg}, e) \textsc{nil}, and f) \ac{cty} populations. The rows of each subplot refer to an FPC (from top to bottom, $\varphi_1$, $\varphi_2$, and $\varphi_3$) while the columns of each subplot refer to weekly, monthly, and weather-dependent distributions of the scores of the corresponding FPC. In contour plots, cold (warm) colors are assigned to negative (positive) values.}
	\label{fig:panel2}
\end{figure*}

While the $f_i\left(t\right)$ of the non-aggregated electric stations belonging to different populations are very different among each other, the daily patterns of the spatially aggregated consumptions, i.e., \textsc{nil} and \textsc{cty}, are very similar. This reflects a rapid loss in the capacity of resolving specific patterns, ascribed to particular contractual characteristics, as soon as the spatial aggregation proceeds towards larger areas. In all the analyzed cases, $\varphi_1\left(t\right)$ is very similar to the corresponding $\mu\left(t\right)$ and the variability that it adds to the average profile does not significantly affect the dynamics of the daily consumption profile, which is mostly determined by the combined effect of $\varphi_2\left(t\right)$ and $\varphi_3\left(t\right)$. The distribution of the scores as a function of the day of the week, the month of the year, and the meteorological conditions are grouped by population and shown in a series of sub-panels in Figure~\ref{fig:panel2}. In the following, the FPCA of each population is briefly discussed in detail.

It is worth mentioning that the discussion in subsections~\ref{sec:res}--\ref{sec:pvg} are referred to representative secondary substations, and that they apply to any substation of the same set, or to same level of spatial aggregation.

\subsubsection{RES}\label{sec:res} The $f_i\left(t\right)$ shows two typical peaks, around 12:00\textsc{pm} and 08:00\textsc{pm}, already observed in literature~\cite{panapakidis2012electricity,yang2013review}. The distribution of the scores of $\varphi_1\left(t\right)$ follows the trend of the consumptions, which are higher than $\mu\left(t\right)$ during the weekends and in the colder months, while lower than $\mu\left(t\right)$ in the milder months, exception made for the month of July which is characterized by consumptions slightly higher than the one of the adjacent months of June and August. The trends suggest a correlation between $\varphi_1\left(t\right)$ and the presence at home, which usually determines higher power loads. Since $\varphi_1\left(t\right)$ explains only $\approx 65\%$ of the variability of the $f_i\left(t\right)$ (see Figure~\ref{fig:fig_1}), and since $\varphi_2\left(t\right)$ and $\varphi_3\left(t\right)$ are mostly related to the dynamics of the consumptions, the analyzed \ac{res} station is subjected to large daily variability. The scores of $\varphi_2\left(t\right)$ and $\varphi_3\left(t\right)$ follow similar distributions. Their combined effect is to increase the consumption levels during the day and to decrease the consumptions levels at night, slightly flattening the consumption patterns and shrinking them towards the central hours of the day. This phenomenon, which strongly occurs during the weekends, can be ascribed to more intense home activities, and to the use of air conditioning systems. The hypothesis is further supported by the trend observed in the monthly distributions of the scores of $\varphi_2\left(t\right)$, which is positively peaked in the summer, reflecting an intensive use of conditioning system in the late evening and a lower presence at home during the day. The monthly distribution of the scores of $\varphi_3\left(t\right)$ is of no trivial interpretation. The contour plots on the meteorological dependencies reveal the existence of a ``comfort zone'' in the temperature interval 15--25 \si{\degreeCelsius} and relative humidity range 20--60\%, where the scores of $\varphi_1\left(t\right)$ and $\varphi_2\left(t\right)$ are negative and the electrical consumptions reach a minimum.

\subsubsection{NRS} The $f_i\left(t\right)$ shows the typical profile of tertiary commercial activities, with high consumption levels extending from the early morning to the late evening~\cite{panapakidis2012electricity,yang2013review}. The distribution of the scores of $\varphi_1\left(t\right)$ follows the consumptions, which are above $\mu\left(t\right)$ during the weekdays, and in general in the colder months of the year, exception made for the month of July, when they reach levels similar to the ones in winter. The observed trend suggests for a correlation between $\varphi_1\left(t\right)$ and the regular commercial activities, which include the use of air conditioning systems. Since $\varphi_1\left(t\right)$ explains more than 80\% of the data variability, $\varphi_2\left(t\right)$ and $\varphi_3\left(t\right)$ play a marginal role. Nevertheless, $\varphi_2\left(t\right)$ has a peculiar shape negatively peaked between 5:00\textsc{am} and 10:00\textsc{am} in the morning. The trend of its daily scores, always negative but Monday and Sunday, and the corresponding monthly trend, negatively peaked in the summer, let suppose a direct correlation between $\varphi_2\left(t\right)$ and the clock-in time, which correspond to an intensive use of air conditioning systems. The effect of $\varphi_3\left(t\right)$ on $\mu\left(t\right)$ is very small when compared to the first two $\varphi_k\left(t\right)$ and of no trivial interpretation. The same ``comfort zone'' observed for the \ac{res} station is found for the analyzed \ac{nrs} station.

\subsubsection{PLT} The $f_i\left(t\right)$ shows a unique profile, with consumptions limited to the dark hours of the day. As expected, the scores of $\varphi_1\left(t\right)$, $\varphi_2\left(t\right)$ and $\varphi_3\left(t\right)$ do not follow any daily trend. However, a monthly trend that resembles a sinusoid is evident for all of them. This trend is typically observed in phenomena that depend on the revolution of the Earth around the Sun, as are the hours of daylight. While $\varphi_1\left(t\right)$ captures the shift of the consumption with respect to $\mu\left(t\right)$, the combined effect of $\varphi_2\left(t\right)$ and $\varphi_3\left(t\right)$ is to shrink (or enlarge) the width of the daily time with zero load. It is worth observing that i) the switch from zero to max load is not a step function, but a ramp with a certain steepness that varies through the year; ii) that the median of the distribution of the scores of $\varphi_2\left(t\right)$ reaches a minimum during the months characterized by longer days; and iii) that the median of the distribution of the scores of $\varphi_3\left(t\right)$ follows a sinusoidal trend with frequency two times the one of $\varphi_2\left(t\right)$. Incidentally, it turns out that the scores of $\varphi_3\left(t\right)$ reach their minimum in the months characterized by the switch to/from the daylight saving time. Therefore, while $\varphi_2\left(t\right)$ can be correlated to the number of daylight hours, which directly affects the overall time over which the public lighting turns on, $\varphi_3\left(t\right)$ can be ascribed to the switching to daylight saving time. As expected, no direct correlation between the principal components and the temperature and the relative humidity is observed.

\subsubsection{PVG}\label{sec:pvg} The $f_i\left(t\right)$ shows a profile that is similar to the one observed in the \ac{nrs} substation, with an additional peculiar cuspid around midday that demonstrates the effect of the self-generated and -consumed electricity. The daily and monthly distributions of the scores of $\varphi_1\left(t\right)$ pretty well resemble the trend observed for the \ac{nrs} substation and thus, following the same reasoning adopted in that case, $\varphi_1\left(t\right)$ can be logically correlated to regular commercial activity. The distribution of the scores of $\varphi_2\left(t\right)$ follows the dynamics of the two peaks of $f_i\left(t\right)$, while the one of $\varphi_3\left(t\right)$ follows the height of the cuspid. While the interpretation of $\varphi_2\left(t\right)$ is not trivial, the monthly distribution of the scores of $\varphi_3\left(t\right)$ are such that the cuspid is less deep during the summer. This could be ascribed to a stronger electric absorption that occurs during the warmer months because of the air conditioning systems, regardless of a larger photovoltaic generation. In fact, it has to be noted that the actual generation installed on the analyzed electric station might be much lower than the registered contractual generation.

\subsubsection{NIL and CTY} The $f_i\left(t\right)$ of the spatially aggregated substations have much in common, with a first peak in the late morning and a second peak in the early evening. The first three $\varphi_k\left(t\right)$ are similar to each other, and their scores follow the same weekly and monthly trends. On the one hand, the median of the weekly distribution of the scores of $\varphi_1\left(t\right)$ is such that the consumptions are above the average level during the weekdays, and below the average level during the weekends, while the scores of $\varphi_2\left(t\right)$ and $\varphi_3\left(t\right)$ are such that the dynamics of the consumption patterns in the early morning and in the evening are mostly affected during the weekends, with a reduction of the morning consumption and an increase of the consumption in the late hours. On the other hand, the monthly distributions of the scores indicate consumptions above the average in cold and hot months, exception made for August, which is characterized by reduced industrial activity and presence, and thus lower consumptions. It is interesting observing the existence of a ``comfort zone'' even in the case of spatially aggregated consumptions, and that this zone is again in the temperature interval 15--25\si{\degreeCelsius} and relative humidity range 20--60\%.

\subsection{Prediction}
\subsubsection{A comparison on a benchmark dataset}
The comparison between the proposed method and some existing work on long-term forecast of electricity load is here presented, using a public domain dataset made available for the EUNITE competition \cite{EUNITE}, consisting in electricity load demand values recorded every two hours in Eastern Slovakia in 1997 and 1998. This dataset has already been used for the long-term forecasting in \cite{Ghelardoni2013}, where data recorded in 1997 are used as training set to predict electricity load demand in 1998. In \cite{Ghelardoni2013}, authors report the performance of several predictive methods using five indices, i.e., the Mean Absolute Error (MAE), the Mean Absolute Percentage Error (MAPE), the Normalized Mean Square Error (NMSE), the Relative  Error Percentage (REP) and the Pearson Product-Moment Correlation Coefficient (PPMCC). We applied the proposed method to this dataset, using data from 1997 for the estimation of the daily function principal components and the prediction models for the correspondent scores. We then predict the electric load demands for 1998 using the first $K=4$ functional principal components, since they explain more than $99\%$ of the variability in the training set. The accuracy of the prediction is measured with the same indices used in \cite{Ghelardoni2013} and the result can be seen in Table \ref{tab:indicesEUNITE}. The long-term predictor based on daily function principal component scores outperforms all the methods reported in Table I in \cite{Ghelardoni2013} for all the considered indices.

\begin{table}[h]
\caption{\label{tab:indicesEUNITE} Performance of the proposed
method in the case of the dataset from EUNITE competition. The FPCA-based prediction outperforms all the methods presented in Table 1 of~\cite{Ghelardoni2013}. Among the many methods presented in~\cite{Ghelardoni2013}, only the results of the best performing one (i.e., SVP+SVB) are here reported.}
\centering
\begin{tabular}{l|r@{.}l|r@{.}l|r@{.}l|r@{.}l|r@{.}l}
\toprule
 & \multicolumn{2}{c|}{MAE} & \multicolumn{2}{c|}{MAPE} &  \multicolumn{2}{c|}{NMSE} & \multicolumn{2}{c|}{REP} &  \multicolumn{2}{c}{PPMCC} \\
\midrule
FPCA & 27&7  & 4&7  & 0&1 & 5&7 &  0&94\\
SVP + SVB & 43&0 & 7&0 & 0&5 & 8&7 & 0&88\\
\bottomrule
\end{tabular}
\end{table}

The final model for the score of the first principal component, which explains more than $93\%$ of the variability in the training set, includes as predictors the month, the day of the week, the day of the month and the interaction between the day of the month and the month (meaning that the coefficient associated to the day of the month changes from month to month).

\subsubsection{Results for electricity consumption in the Milan area}
The FPCA method has been used to predict the power consumption profiles of each substation. As the aim of the present paper is to show the potential benefits in using FPCA, efforts have not been put in optimizing the selection of the training set, of the prediction model, and of the predictors. The forecast was carried out on a long-term basis. The period 2014--2015 was used to predict the hourly power profiles for the whole 2016 for each selected substation.

\begin{table*}[t]
	\centering
    \caption{Yearly MAPEs (second column) and monthly energy percentage error for each representative substation in each population.}\label{tab:tab_2}
    \includegraphics[width=\textwidth]{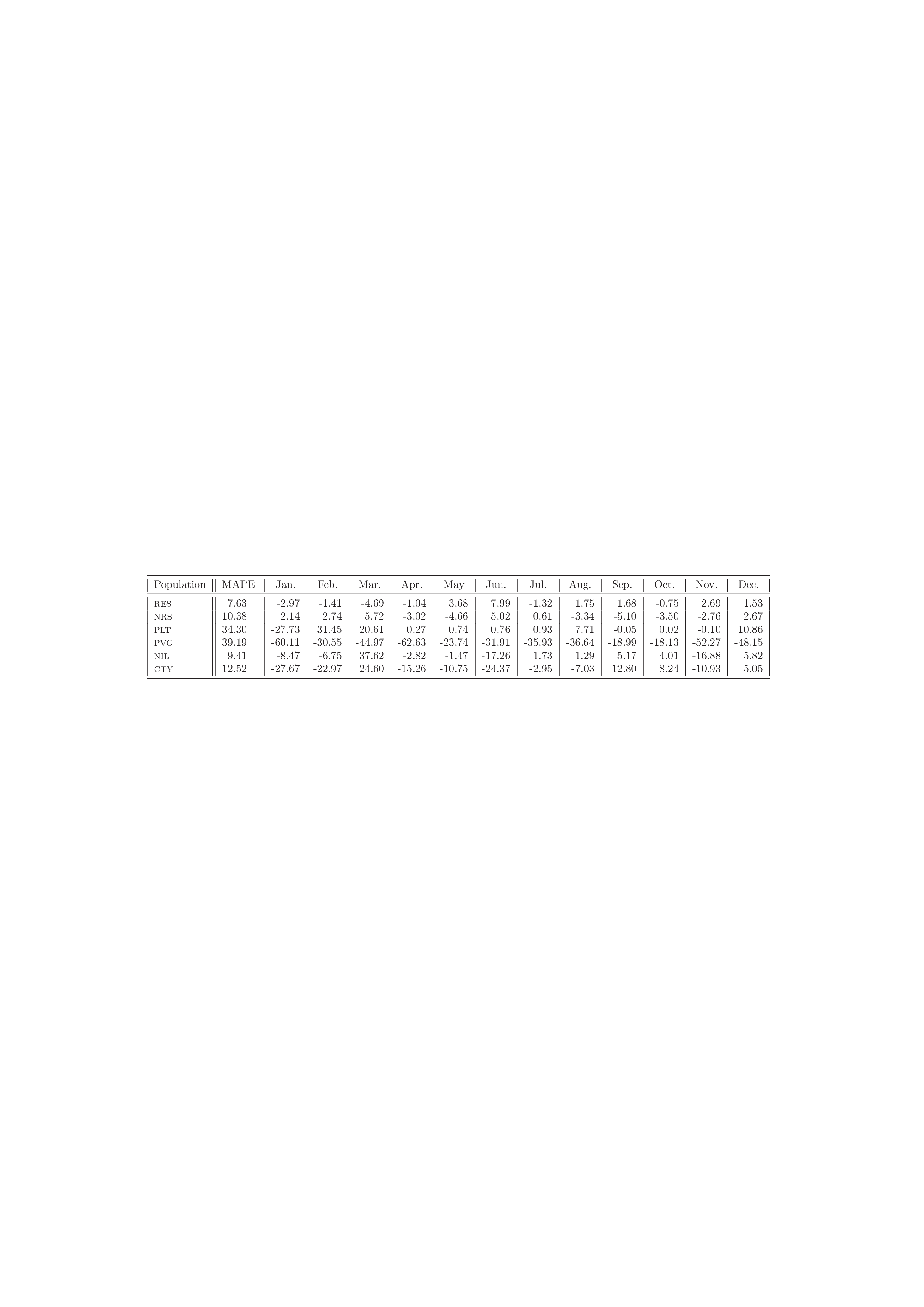}
\end{table*}
\begin{figure*}[t]
	\centering
    \includegraphics[width=.95\textwidth]{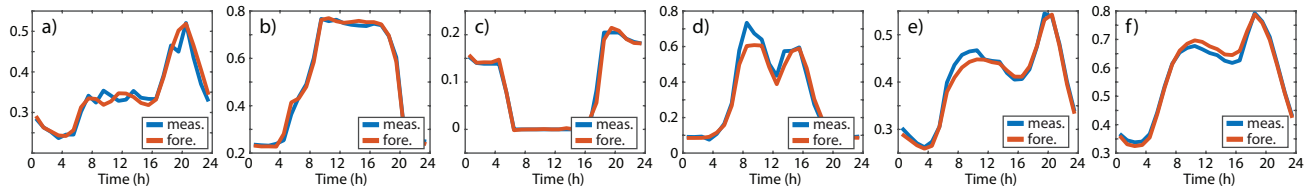}
    	\caption{Predicted (red) and actual (blue) daily load curves of the day with lowest daily MAPE of each representative substation in each population: (a) \ac{res} on 04/08/2016, (b) \ac{nrs} on 03/11/2016, (c) \ac{plt} on 07/23/2016, (d) \ac{pvg} on 06/17/2016, (e) \textsc{nil} on 04/10/2016, and (f) \ac{cty} on 05/21/2016.}
	\label{fig:panel3}
\end{figure*}

The predicted and the actual daily load curve of a given day is shown in \mbox{Figure~\ref{fig:panel3}a--f} per each representative element of each population studied in this work. Each plot shows the day with the lowest daily MAPE. The corresponding yearly MAPEs are shown in the second column of  Table~\ref{tab:tab_2}. The prediction model predicts and reproduces the actual load curves of the population \ac{res} with good accuracy. The errors for \textsc{plt} and \textsc{pvg} recommend that further analysis and customization should be dedicated to better predict these kinds of substations. It also worth noting that the prediction occurred once for the whole year. The error on the prediction does not decrease with the spatial aggregation, i.e., with the loss of resolution on contractual characteristics. This goes against what $\vartheta$ would suggest (see Figure~\ref{fig:fig_1}), which increases with the spatial aggregation, and might be consequence of the choice of the linear model as a good compromise between complexity and accuracy.

The percentage error on the average monthly energy is reported in Table~\ref{tab:tab_2} per month per population. The monthly energy estimates \ac{res} and \ac{nrs} have $\varepsilon_{\%}< 10\%$, while the error on \ac{plt}, \ac{pvg}, \textsc{nil} and \ac{cty} is larger and depends on the month of the year.

\section{Conclusion and Perspectives}
This work demonstrates that the FPCA is a versatile technique that can be exploited to study the electric consumption patterns at various level of spatial aggregation. The FPCA exploits unique properties that allows to establish a compromise between complexity and amount of variability explained by decomposing, or predicting, the daily patterns on the basis of a selected limited number of $\varphi_k\left(t\right)$. To demonstrate the capabilities of the method, the FPCA was applied to three years of unique and sensitive historical data of electric consumption at the distribution level. The first three functional principal components were found sufficient to explain more than 80\% of the variability of the data per electric station type and/or spatial aggregations, while the correlation between the principal components and the exogenous causes was rapidly lost after the first two components. The first two principal components were found strongly correlated with the calendar periodicity and the weather conditions, the latter allowing to identify a ``comfort region'' where the consumptions reach a minimum. A linear prediction algorithm, based on the FPCA decomposition method and chosen to minimize the Akaike Information Criterion (AIC), demonstrated that FPCA-based linear models have interesting capabilities in predicting time series both in the short- and in the mid/long-term. This opens the doors to further investigations that will aim at understanding advantages and disadvantages of FPCA-based linear models with respect to other more common models, such as ARIMA and Neural Networks.

In this work, for the sake of simplicity a linear model is used to predict future scores. Also, the model is selected in an automatic way with no fine-tuning for the different types of substation. More complex models can of course be identified for the scores prediction and this can be scope for future work. Moreover, we expect that different predictive models will be needed for the different types of substations and this model building can be informed by the analysis we presented. Additional features can also be introduced, in particular meteorological variables.

To conclude, this work assesses the FPCA as an innovative and powerful method to investigate and predict the electric consumption patterns at any spatial aggregation level and opens the doors to further studies that aim at optimizing the algorithm for predictive purposes.

\IEEEtriggeratref{23}

\bibliographystyle{IEEEtran}

\end{document}